\documentclass[a4,11pt]{article}
\usepackage{graphicx}

\title{Cellular Automata Networks}

\author{Xin-She Yang\thanks{Corresponding Author}   $\;\;$ and Young Z. L. Yang  \\
Department of Engineering, University of Cambridge \\
Trumpington Street, Cambridge CB2 1PZ, UK  }

\date{}
\begin{document}
\maketitle

\begin{abstract}
{\large
A small-world cellular automaton network has been formulated to
simulate the long-range interactions of complex networks using
unconventional computing methods in this paper. Conventional
cellular automata use local updating rules. The new type of cellular
automata networks uses local rules with a fraction of long-range
shortcuts derived from the properties of small-world networks.
Simulations show that the self-organized criticality emerges
naturally in the system for a given probability of shortcuts and
transition occurs as the probability increases to some critical
value indicating the small-world behaviour of the complex automata
networks. Pattern formation of cellular automata networks and the
comparison with equation-based reaction-diffusion systems are also discussed.} \\

\index{cellular automata} \index{automata network}


\end{abstract}

\noindent {\bf Citation detail:} X. S. Yang and Y. Z. L. Yang,
Cellular automata networks, {\it Proceedings of Unconventional Computing 2007} (Eds. A. Adamatzky,
L. Bull, B.De Lacy Costello, S. Stepney, C. Teuscher), Luniver Press, pp. 280-302 (2007).

\section{Introduction}

Theory and computation about complex networks such as the bacterial
colonies, interacting ecological species, and the spreading of
computer virus over the Internet are becoming very promising and
they may have important applications in a wide range of areas. The
proper modelling of these networks is a challenging task and the
studies in this area are still at very early stage. However, various
techniques and applications have been investigated, especially in
the area of computational logic, the Internet network, and
application of bio-inspired algorithms [1-7]. Since the pioneer work
of Watts and Strogatz on small-world networks, a lot of interesting
studies on the theory and application of small-world networks
[7-9,12-18] have been initiated. More recently, the automata
networks have been developed by Tomassini and his colleagues
\cite{Tomassini,Tomas1} to study the automata network in noise
environment. Their study shows that small-world automata networks
are less affected by random noise. The properties of complex
networks such as population interactions, Internet servers, forest
fires, ecological species and financial transactions are mainly
determined by the way of connections between the vertices or
occupied sites. \index{small-world networks}

Network modelling and formulations are essentially discrete in the
sense that they deal with discrete interactions among discrete nodes
of networks because almost all the formulations are in terms of the
degrees of clustering, connectivity, average nodal distance and
other countable degrees of freedom. Therefore, they do not work well
for interactions over continuous networks and media. In the later
case, modelling and computations are usually carried out in terms
partial differential equations (PDEs), however, almost all PDEs
(except those with integral boundary conditions) are local equations
because the derivatives and the dependent variable are all evaluated
at concurrent locations. For example, the 3-D reaction-diffusion
equation \begin{equation} u_t=D \nabla^2 u + r(u;x,y,z,t),
\end{equation} describes the variation of $u(x,y,z,t)$ such as temperature
and concentration with spatial coordinates $(x,y,z)$ and time $t$.
While the diffusion coefficient $D$ can be constant, but the
reaction rate $r$ may depends on $u$ and location $(x,y,z)$ as well
as time $t$. This equation is local because $u$, $u_t$, $r$ and
$\nabla^2 u$ are all evaluated at the same point $(x,y,z)$ at any
given time $t$. Now if we introduce some long-distance shortcuts
(e.g., a computer virus can spread from one computer to another
computer over a long-distance, not necessary local computers), then
the reaction rate can have a nonlocal influence in a similar manner.
We can now modify the above equation as \begin{equation} u_t = D
\nabla^2 u + q\Big(u(x,y,z,t), u(x_*,y_*,z_*,t);x,y,z,t\Big),
\end{equation} where $q$ depends on the local point $(x,y,z)$ and
another point $(x_*,y_*,z_*)$ far away. Obviously, $q$ can be any
function form. As a simple example in the 1-D case: $u_t=D u_{xx} +
u(x,t)*[1-u(x,t)]+\beta u(x-s,t)$, where $s$ is simply a shift and
$\beta \in [0,1]$ is a constant. This equation is nonlocal since the
reaction rate depends the values of $u$ at both $x$ and $x-s$ at
$t$. This simple extension makes it difficult to find analytical
solutions. Even numerical solutions are not easy to find because the
standard numerical methods do not necessarily converge due to the
extra nonlocal term. This paper will investigate this aspect in
detail using unconventional solution methods such as small-world
cellular automata networks.

The present work aims to develop a new type of small-world cellular
automata by combining local updating rules with a probability of
long-range shortcuts to simulate the interactions and behaviour of a
complex system. By using a small fraction of sparse long-range
shortcut interactions together with the local interactions, we can
simulate the evolution of complex networks. Self-organized
criticality will be tested based on the results from the cellular
automata. The important implications in the modelling and
applications will also be discussed.

\section{Small-World Networks}

Small-world networks are a special class of networks with a high
degree of local clustering as well as  a small average distance, and
this small-world phenomenon can be achieved by adding randomly only
a small fraction of the long-range connections, and some common
networks such as power grids, financial networks and neural networks
behave like small-world networks \cite{moukarzel,newman2000}. The
application of small-world networks into the modelling of infection
occurring locally and at a distance was first carried out by Boots
and Sasaki \cite{boots} with some interesting results. The dynamic
features such as spreading and response of an influence over a
network have also been investigated in recent studies
\cite{newman2000} by using shortest paths in system with sparse
long-range connections in the frame work of small-world models.  The
influence propagates from the infected site to all uninfected sites
connected to it via a link at each time step, whenever a long-range
connection or shortcut is met; the influence is newly activated at
the other end of the shortcut so as to simulate long-range sparkling
effect. These phenomena have successfully been studied by Newman and
Watts model \cite{newman99} and Moukarzel \cite{moukarzel}. Their
models are linear in the sense that the governing equation is linear
and the response is immediate as there is no time delay in their
models \cite{yang01}. More recently, one of the most interesting
studies has been carried out by De Arcaneglis and Herrmann
\cite{arc02} using the classic height model on a lattice, which
implied the self-organized criticality in the small-world system
concerned. \index{self-organized criticality}

On the other hand, cellular automata have been used to simulate many
processes such as lattice gas, fluid flow, reaction-diffusion and
complex systems \cite{weimar,wolfram94,yang01,yang02,yang03} in
terms of interaction rules rather than the conventional partial
differential equations. Compared to the equation-based models,
simulations in term of cellular automata are more stable due to
their finite states and local interacting  rules \cite{weimar}. In
fact, in most cases, the PDE models are equivalent to rule-based
cellular automata if the local rules can be derived from the
corresponding PDE models \cite{wolfram94,guinot}, and thus both PDE
models and CA rules can simulate the same process \cite{yang03}.
However, we will show that cellular automata networks are a better
approach for solving nonlocal equation-based models.

The rest of the present paper will focus on: 1) to formulate a
cellular automaton network on a 2-D lattice grid with sparse
long-range shortcuts; 2) to simulate the transition and complexity
concerning small-world nonlocal interactions; 3) to test the
self-organized criticality of the constructed network systems; 4) to
find the characteristics of any possible transition.

\section{Cellular Automata Networks}

\index{cellular automata network}

Earlier studies on cellular automata use local rules updating the
state of each cell and the influence is local. That is to say, the
state at the next time step is determined by the states of the
present cell concerned and those of its immediate surrounding
neighbour cells. Even the simple rules can produce complex patterns
\cite{wolfram94}.  The rule and its locality determine the
characteristics of the cellular automata. In fact, we do not have to
restrict that the rules must be local, and in general the influence
can be either local or nonlocal. Thus, we can assume the rules of
cellular automata can be either local or nonlocal or even global.
The state of a cell can be determined by $m$ cells consisting of
$m_i$ immediate neighbour cells and $m_o=m-m_i$ other cells at
longer distance. In the case of local rules only, $m=m_i$ and
$m_o=0$. If $m_o \ne 0$, then the rules are nonlocal. If $m$ is the
same order of the total cells $N \times N$ of the cellular
automaton, then rules are global. Nonlocal interactions rule for
lattice-gas system was first developed by Appert and Zaleski in the
discussion of a new momentum-conserving lattice-gas model allowing
the particles exchange momentum between distant sites \cite{appert}.
Some properties of local and nonlocal site exchange deterministic
cellular automata were investigated by researchers
\cite{Tomas1,boc}. As the nonlocal rules are different from the
local rules, it is naturally expected that the nonlocal rules may
lead to different behavior from conventional local rule-based
cellular automata. Furthermore, self-organized criticality has been
found in many systems in nature pioneered by Bak and his colleagues
\cite{bak87,bak96}. One can expect that there may be cases when
self-organized criticality, cellular automata, and small-world
phenomena can occur at the same time. More specifically, if a
finite-state cellular automaton with a small-fraction of long-range
shortcuts is formulated, a natural question is: Do the
self-organized criticality exist in the small-world cellular
automaton? Is there any transition in the system?

\subsection{Local Cellular Automata}

A cellular automaton is a finite-state machine defined on a regular
lattice in $d$-dimensional case, and  the state of a cell is
determined by the current state of the cell and the states of its
neighbour cells \cite{weimar,wolfram94}. For simplicity, we use 2-D
in our discussions. A state $\phi_{i,j}$ of a cell $(i,j)$ at time
step $n+1$ can be written in terms of the previous states
\begin{equation} \phi^{n+1}_{i,j}= \sum_{k,l=-r}^{r} c_{k,l}
\phi^n_{i+k,j+l}, \;\; i,j=1,2,...,N
\end{equation} where summation is over the $4r(r+1)$ Moore
neighbourhood cells.
In the $d$-dimensional case, there are $(2r+1)^d-1$ Moore
neighbourhood cells.  $N$ is the size of the 2-D automaton, and
$c_{k,l}$ are the coefficients. For the simplest and well-known 2-D
Conway's game of life $c_{k,l}=1$ for 8 neighbour cells ($r=1$). Now
let us introduce some nonlocal influence from some sparse long-range
cells (see Fig. 1) by combining  small-world long-range shortcuts
and conventional cellular automata to form a new type of  cellular
automata networks.

\subsection{Small-World Automata Networks}

For simplicity, we define a small-world cellular automaton network
as a local cellular automaton with an additional fraction or
probability $p$ of sparse long-range nonlocal shortcuts (see Fig.
1). For $m_i=4 r_m +1 $ immediate Neumann neighbours and $m_o=2 r_o
$ nonlocal cells, the updating rule for a cell becomes
\begin{equation} \phi^{n+1}_{i,j}= \sum_{k,l=-r_m}^{r_m} c_{k,l}
\phi^n_{i+k,j+l} + \delta (p) \sum_{s,q=-r_o}^{r_o} c_{s,q}
\phi^n_{i+d_i+s,j+d_j+q}, \label{equ-rule} \end{equation} where
$\delta (p)$ is a control parameter that can turn the long-range
cells on ($\delta=1$) or off ($\delta=0$) depending on the
probability $p$.  \index{small-world automata}

The probability $p$ is the fractions of long-range shortcuts in the
total of every possible combinations. For $N \times N$ cells, there
are $N^2 (N^2-1)$ possible connections. The simplest form of
$\delta$ can be written as $\delta(p)=p H(p-p_0)$ where $H$ is a
Heaviside function. $p_0$ is a critical probability, and $p_0$ can
be taken as to be zero in most simulations in this paper. The
updating rules are additive and thus form a subclass of special
rules. We can extend the above updating rules to a generalized form,
but we are only interested in additive rules here because they may
have interesting properties and can easily be transformed to
differential equations. In addition, the neighourhood can be either
extended Moore neighourhood or Neumann neighbourhood. For Moore
neighbourhood, $m_i=4r_m (r_m+1)$ and $m_o=4 r_o$. Our numerical
experiments seem to indicate that Moore neighbourhood is more
sensitive for avalanche and Neumann neighbourhood is more stable for
pattern formation. A simple case for a small-world cellular
automaton in 2-D case is $r_m=1$ and $r_0=5$ (or any $r_0>r_m$) so
that it has 5 immediate Neumann neighbour cells and 2 shortcuts.

The distance between the nonlocal cells to cell $(i,j)$ can be
defined as \begin{equation} S=\sqrt{s_i^2+s_j^2}. \end{equation} The
nonlocality requires \begin{equation} \min(|s_i|, |s_j|)>r_m.
\end{equation} The nonlocal influence can also be introduced in other ways.
Alternatively, we can use the conventional local rule-based cellular
automaton and adding the long-distance shortcuts between some cells
in a random manner. The probability $p$ of the long-range shortcuts
in all the possible connections is usually very small. Under certain
conditions, these two formulations are equivalent. More generally, a
finite-state cellular automaton with a transition rule $G=[g_{ij}]
\;\;(i,j=1,2,...,N)$ from one state $\Phi^n=[\phi_{ij}^n]
\;\;(i,j=1,2,...,N)$ at time level $n$ to a new state
$\Phi^{n+1}=[\phi^{n+1}_{ij}] \;\;(i,j=1,2,...,N)$ at time level
$n+1$ can be written as \begin{equation} G: \Phi^n \mapsto
\Phi^{n+1}, \; g_{ij}: \phi^{n}_{ij} \mapsto \phi^{n+1}_{ij},
\end{equation} where $g_{ij}$ takes the same form as equation
(\ref{equ-rule}) for small-world cellular automata.

\begin{figure}
\centerline{\includegraphics[width=3.5in,height=2in]{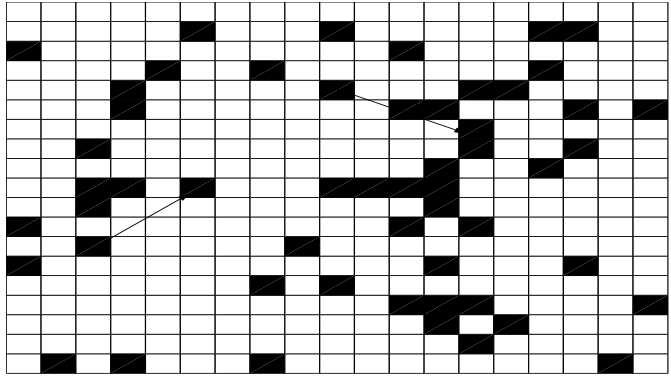}}
\caption{ A cellular automaton network with a probability $p$ of
long-range shortcuts }
\end{figure}
The state of each cell can be taken to be discrete or continuous.
From simplicity, we use $n_v$-valued discrete system and for most of
the simulations in the rest of the paper, we use $n_v=2$ (thus, each
cell can only be $0$ or $1$) for self-criticality testing, and
$n_v=1024$ for pattern formation. Other numbers of states can be
used to meet the need of higher accuracies.

\section{Simulations and Results}

By using the small-world cellular automaton formulated in the
previous section, a large number of computer simulations have been
carried out in order to find the statistic characteristics of the
complex patterns and behaviour arising from cellular automata
networks with different probabilities $p$ of long-range shortcuts.
Numerical simulations are carried out on an $N \times N$ lattice in
2-D setting, and usually, $N \ge 40$, or up to $5000$. Different
simulations with different lattice size are compared to ensure the
simulated results are independent of the lattice size and time
steps. In the rest of the paper, we present some results concerning
the features of transition and self-organized criticality of
small-world cellular automata.

\subsection{Self-organized criticality}

For a lattice size of $N=2000 \times 2000$ with a fixed $p$, a
single cell is randomly selected and perturbed by flipping its state
in order to simulate an event of avalanche in 2-D automata networks
with the standard Moore neighbourhood and Game-of-life updating
rules, but a probability $p$ is used to add long-range shortcuts to
the cellular automaton. A shortcut forces the two connecting cells
having the same state. Figure 2 shows the avalanche size
distribution for two different values of $p=0.05$ and $p=0.2$,
respectively. The avalanche size is defined as the number of cells
affected by any single flipping perturbation.

In the double logarithmic plot, the data follows two straight lines.
It is clearly seen that there exists a power law in the
distribution, and the gradient of the straight line is the exponent
of the power-law distribution. A least-square fitting of $N \propto
s^{-\gamma}$, leads to the exponents of $\gamma=1.06 \pm 0.04$ for
$p=0.05$ and $\gamma=1.40 \pm 0.05$ for $p=0.2$. Although a
power-law distribution does not necessarily mean the self-organized
criticality. Self-organized criticality has been observed in other
systems [1-4,26]. The pattern formed in the system is quasi-stable
and a little perturbation to the equilibrium state usually causes
avalanche-like readjustment in the system imply the self-organized
criticality in the evolution of complex patterns of the cellular
automaton. This is the first time of its kind by using computer
simulations to demonstrate the feature of self-organized criticality
on a {\it cellular automaton network}. We can also see in Figure 2
that different probabilities $p$ will lead to different values of
exponents. The higher the probability, the steeper the slope.

\begin{figure}
\centerline{\includegraphics[width=3in,height=2.25in]{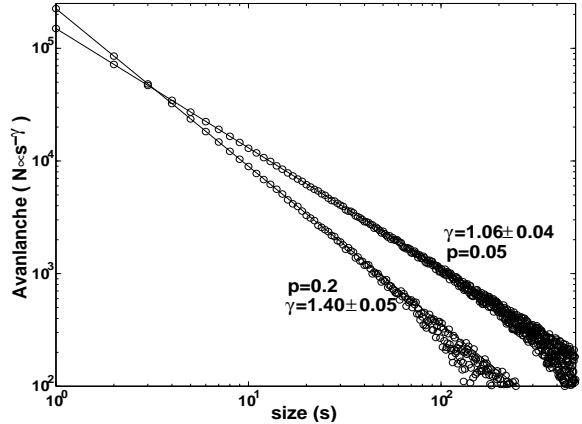}}
\caption{ Avalanche size distribution for a small-world cellular
automaton. }
\end{figure}

\subsection{Transition of small-world systems}

For a fixed grid of $2000 \times 2000$ cells, we can vary the
probability $p$ to see what can happen. For a single event of
flipping state, the fraction of population affected is plotted
versus $p$ in a semi-logarithmic plot as shown in Figure 3 where the
fraction of population is defined as the number ($N_a$) of cells
affected among the whole population $N^2$, that is $N_a/N^2$. The
sharp increase of the fraction versus the probability $p$ indicates
a transition in the properties of cellular automata networks. For a
very small probability $p<0.004$, the influence of the event mainly
behaves in a similar way as the conventional local cellular
automata. As the probability increases, a transition occurs at about
$p=0.01$. For $p>0.05$, any event will affect the whole population.
This feature of transition is consistent with the typical
small-world networks \cite{newman99,yang02}.

\begin{figure}
\centerline{\includegraphics[width=3in,height=2.5in]{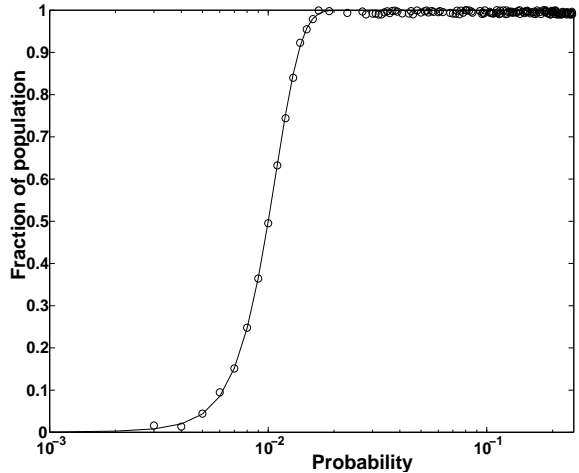}}
\caption{ Transition of influence with different probabilities $p$
of long-range shortcuts.}
\end{figure}

Comparing with the local rule-based cellular automata, the
transition in small-world cellular automata is an interesting
feature. Without such shortcuts, there was no transition observed in
the simulations. However, self-organized criticality was still
observed in finite-state cellular automata \cite{adami,yang02}
without transition. In the present case, both self-organized
criticality and transition emerge naturally. Thus, the transition in
cellular automata networks suggests that this transition may be the
result of nonlocal interactions by long-range shortcuts.

This feature of transition may have important implications when
applied to the modelling of real-world phenomena such as the
Internet and social networks. For a system with few or no long-range
interactions, there is no noticeable change in its behavior in
transition. However, as the long-range shortcuts or interacting
components increase a little bit more, say to $p=0.01$, then a
transition may occur and thus any event can affect a large fraction
of the whole population. For example, to increase the speed of
finding information on the Internet, a small fraction of long-range
shortcuts in terms of website portals and search engine (e.g.,
Google) and high-capacity/bandwidth connections could significantly
increase the performance of the system concerned. In addition, the
self-organized criticality can also imply some interesting
properties of the Internet and other small-world networks, and these
could serve as some topics for further research.

\subsection{Nonlocal Partial Differential Equations}

\index{partial differential equation}

The evolution of a system can usually be described by two major
ways: rule-based systems and equation-based systems. The rule-based
systems are typically discrete and use local rules such as cellular
automata or finite difference system. As discussed by many
researchers \cite{weimar,yang03}, the finite difference systems are
equivalent to cellular automata if the updating rules for the
cellular automata are derived directly from their equation-based
counterpart. On the other hand, the equation-based systems are
typically continuous and they are often written as partial
differential equations. Sometimes, the same system can described
using these two different ways. However, there is no universal
relationship between a rule-based system and an equation-based
system \cite{yang03}. Given differential equation, it is possible to
construct a rule-based cellular automaton by discretizing the
differential equations, but it is far more complicated to formulate
a system of partial differential equations for a given cellular
automaton. For example, the following 2-D partial differential
equation for nonlinear pattern formation for $u(x,y,t)$
\begin{equation} \frac{\partial u}{\partial t}=D (\frac{\partial^2
u}{\partial x^2}+\frac{\partial^2 u}{\partial y^2} )+\gamma
f(u,x,y,t), \label{equ-1} \end{equation} can always be written as an
equivalent cellular automaton if the local rules of cellular
automaton are obtained from the PDE. Conversely, a local cellular
automaton can lead to a local system of partial differential
equation (PDE), if the construction is possible \cite{yang03}. A
local PDE can generally be written as \begin{equation} {\cal F}(u,
u_x, u_y,...,x,y,t)=0. \end{equation}

A nonlocal PDE can be written as \begin{equation} {\cal F}(u, u_x,
u_y,...,x,y,t, x+S(x,y,p),y+S(x,y,p))=0, \end{equation} where
$S(x,y,p)$ is the the averaged distance of long-range shortcuts and
$p$ is the probability of nonlocal long-range shortcuts. In order to
show what a nonlocal equation means, we modify the above equation
for pattern formation as \begin{equation}  u_t=D \nabla^2
u(x,y,S,p,t)+\gamma u(x,y,S,p,t)[1-u(x,y,S,p,t)]. \end{equation}
This nonlocal equation is far more complicated than equation (8).
For the proposed cellular automaton networks, a system of nonlocal
partial differential equations will be derived, though the explicit
form of a generic form is very difficult to obtain and this requires
further research.
\begin{figure}
\centerline{\includegraphics[width=3in,height=2in]{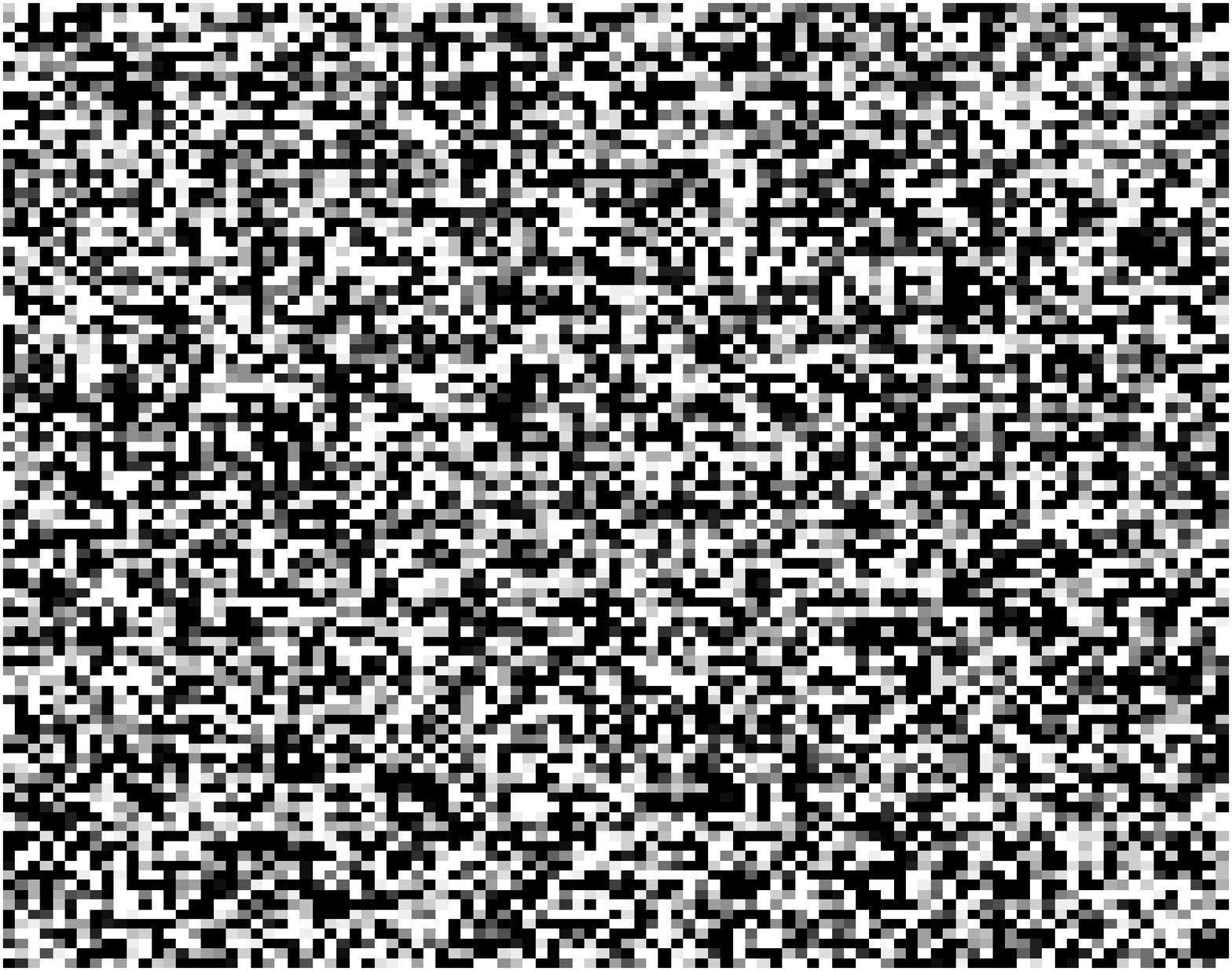}
\includegraphics[width=3in,height=2in]{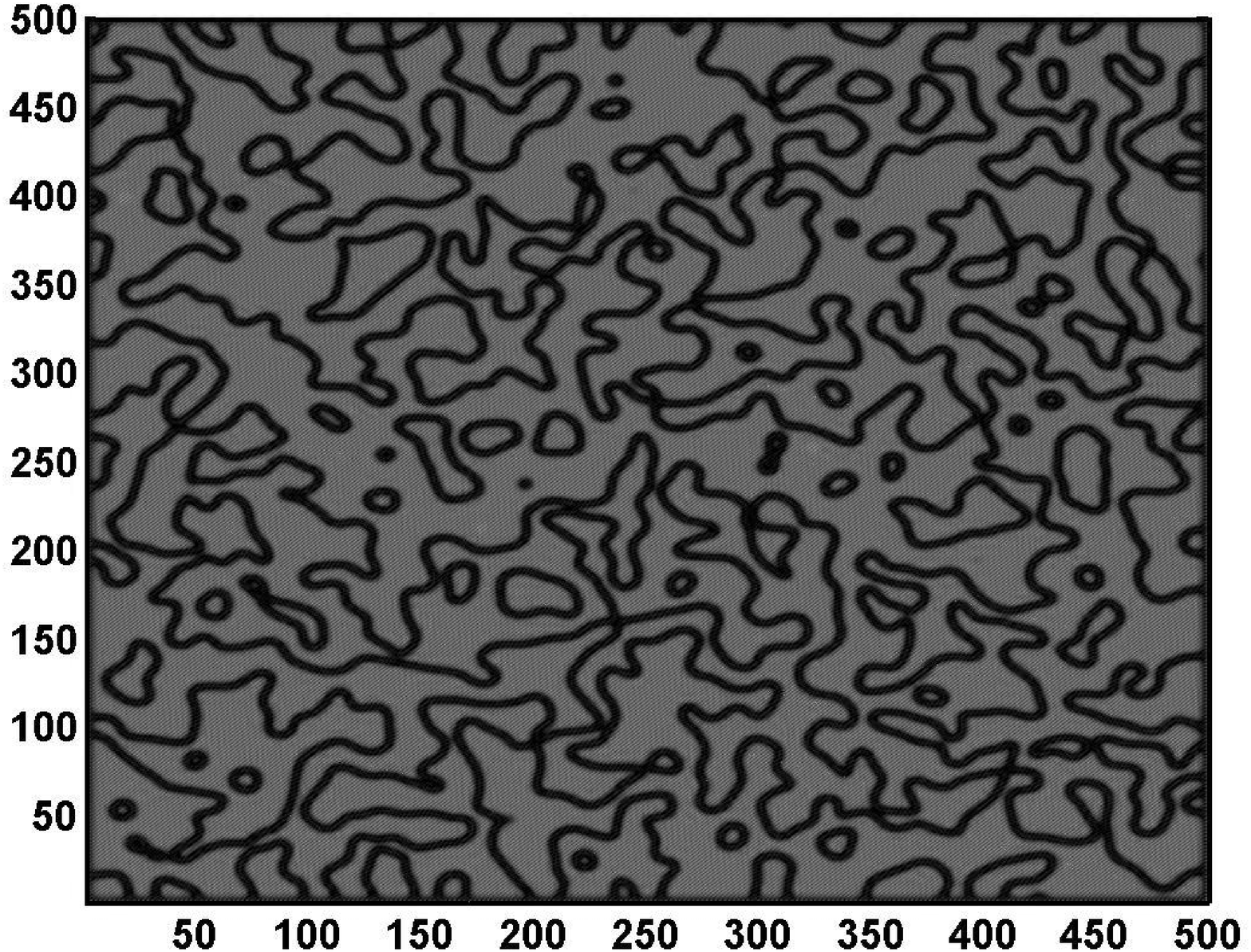}}
\caption{ a) Initial random configuration on a $500 \times 500$ grid
at $t=0$; b) 2-D Pattern formation and distribution displayed at
$t=100$. }
\end{figure}
\subsection{Pattern Formation}

Even for simple nonlinear partial differential equations, complex
pattern formation can arise naturally from initially random states.
For example, the following nonlinear partial differential equation
for pattern formation $u(x,y,t)$ \begin{equation} \frac{\partial
u}{\partial t}=D (\frac{\partial^2 u}{\partial x^2}+\frac{\partial^2
u}{\partial y^2} )+\gamma u(1-u) + \beta u(x-S), \label{equ-2}
\end{equation} can be discretized using central finite difference
scheme in space  with $\Delta x=\Delta y=\Delta t=1$. Then, it is
equivalent to \begin{equation} u^{n+1}_{i,j}=\sum_{k,l=-r}^{r}
a_{k,l} u^n_{i+k,j+l}+\gamma u^n_{i,j}(1-u^n_{i,j})+ \beta
u^n_{i-S,j}, \end{equation} where $r=1$, $a_{0,0}=1-4D,
a_{-1,0}=a_{+1,0}=a_{0,-1}=a_{0,+1}=D$. It is a cellular automaton
with the standard Neumann neighbourhood for this PDE. \index{pattern
formation}

The formed patterns and their distribution resulting from the system
on a $500 \times 500 $ grid are shown in Fig. 4 where $D=0.2$,
$S=5$, $\gamma=0.5$ and $\beta=0.01$. We can see that stable
patterns can be formed from initial random states.

\begin{figure}
\centerline{\includegraphics[width=1.5in,height=1.5in]{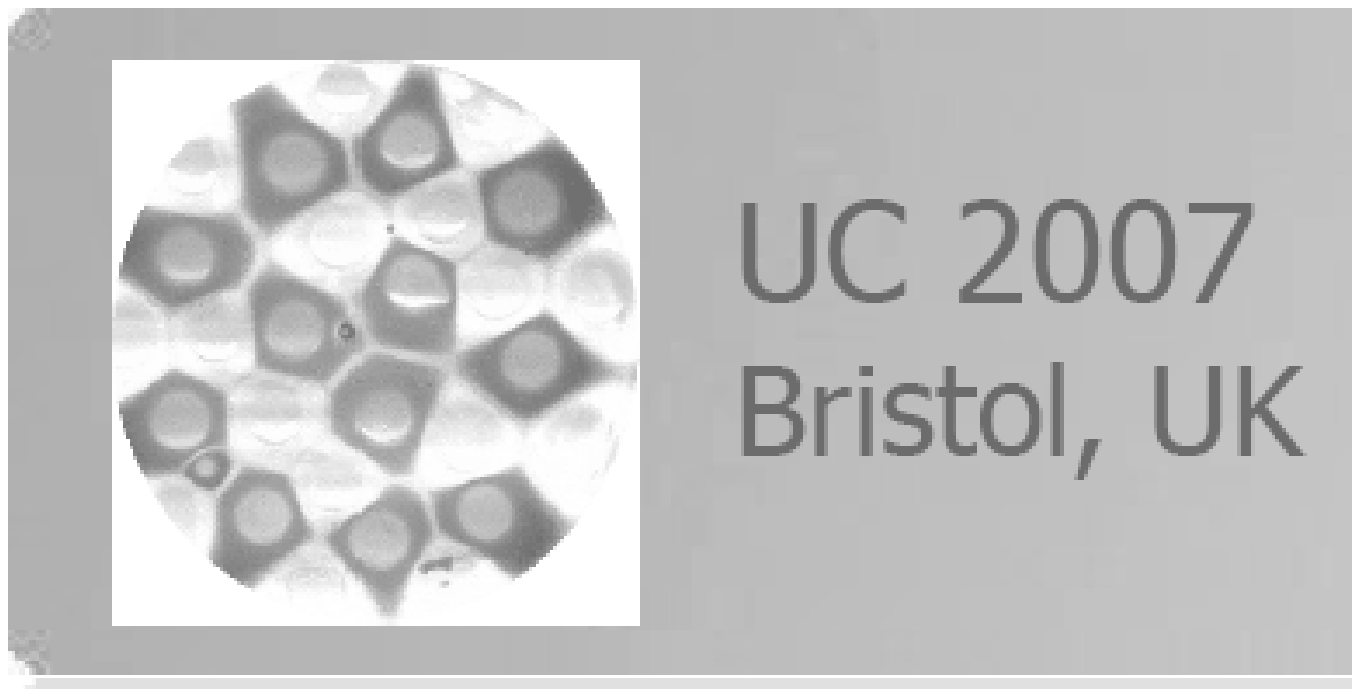}
\includegraphics[width=1.5in,height=1.5in]{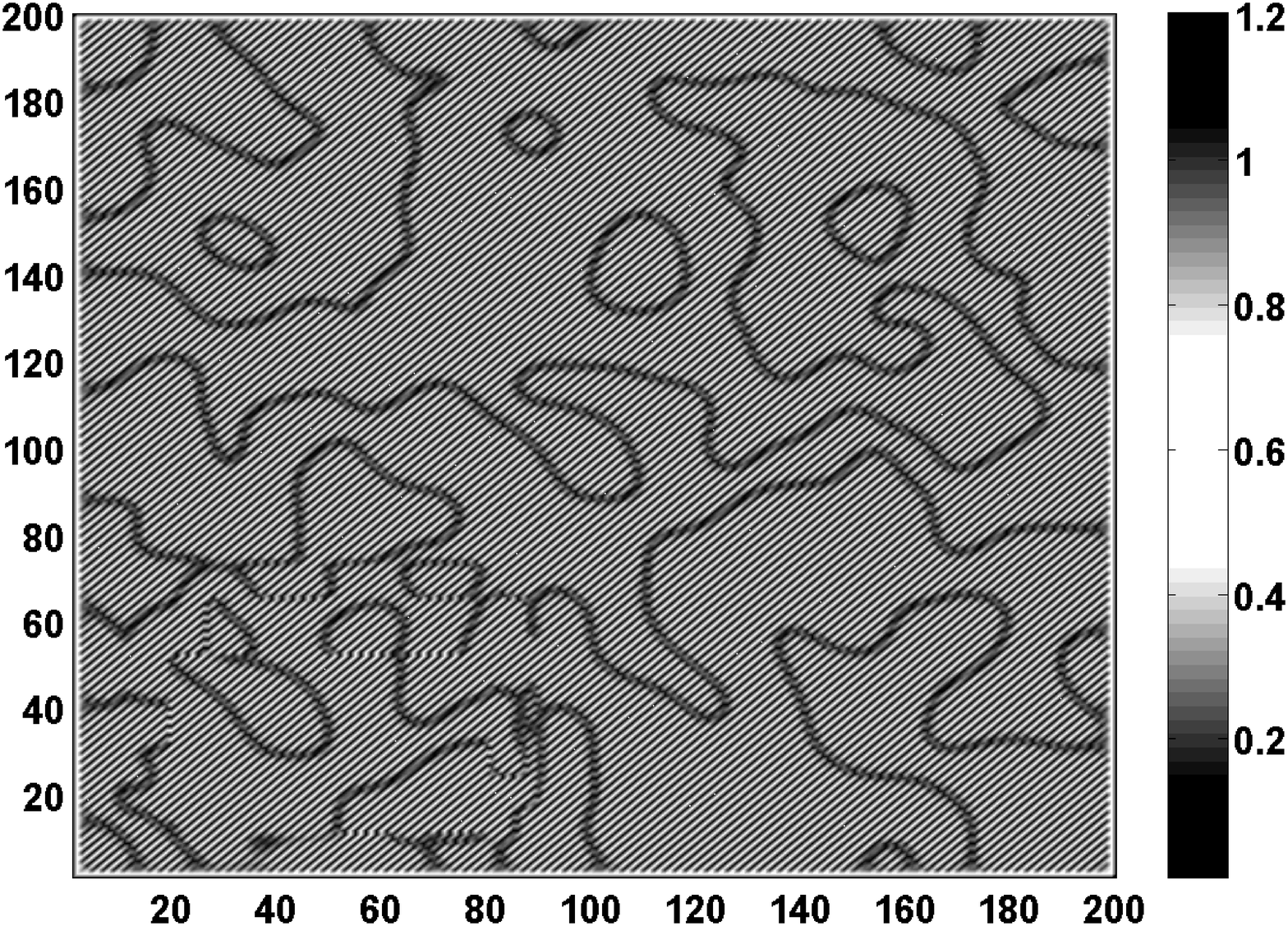}
\includegraphics[width=1.5in,height=1.5in]{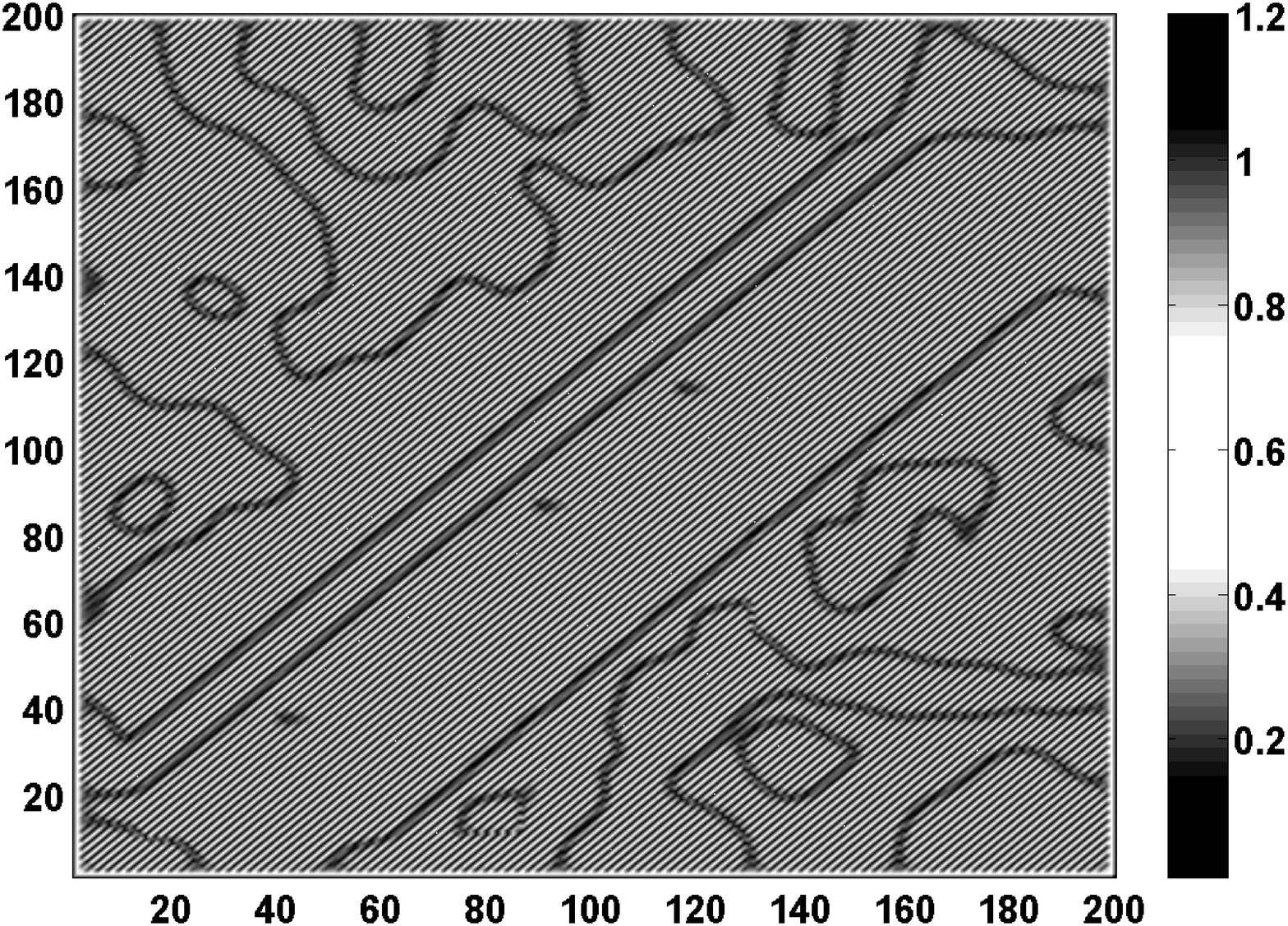}}
\centerline{(a) \hspace{1.5in} (b) \hspace{1.5in} (c)}

\caption{ a) The UC2007 logo has been mapped onto a $200 \times 200$
grid as the initial condition at $t=0$; b) 2-D pattern distribution
with a short-cut probability $p=0.0025$ and $S=5$ (or $\beta=0.05$)
displayed at $t=200$; c) 2-D pattern with a short-cut probability
$p=0.01$ and $S=50$ (or $\beta=0.2$) at $t=200$. }
\end{figure}

The formed patterns using the Neumann neighbourhood ($r=1$) are very
stable and are almost independent of the initial conditions. In
fact, the initial state does not matter and the only requirement for
the initial state is some degree of randomness. If we run the same
program using a photograph or the UC2007 conference logo, similar
patterns can also form naturally as shown in Figure 5. In our
simulations, we have used $D=0.2$ and $\gamma=0.5$. Other parameters
$S$ and $\beta$ can vary. For the case shown in Fig. 5b, $S=5$ and
$\beta=0.05$, while $S=50$ and $\beta=0.2$ are used in Fig. 5c. This
means that the initial state does not affect the characteristics of
pattern formation and this is consistent with the stability analysis
\cite{yang03}.

\section{Discussions}

Small-world cellular automata networks have been formulated to
simulate the interactions and behaviour of multi-agent systems and
small-world complex networks with long-range shortcuts. Simulations
show that power-law distribution emerges for a fixed probability of
long-range shortcuts, which implies self-organized criticality in
the avalanche and evolving complex patterns. For a given size of
cellular grid, the increase of the probability of long-range
shortcuts leads to a transition, and in this case, a single even can
affect a large fraction of the whole population. In this sense, the
characteristics of small-world cellular automota are very different
from the conventional locally interacting cellular automata.

The nonlocal rule-based network systems in terms of cellular
automata can have other complicated features such as its
classifications compared with the conventional automata and its
relationship its partial differential equations. In addition,
cellular automota networks could provide a new avenue for efficient
unconventional computing for simulating complex systems with many
open questions such as the relationship between cellular automata
networks and nonlocal PDEs, and the potential implication on the
parallelism of these algorithms. These are open problems to be
investigated in the future research.


\begin{thebibliography}{A}

\bibitem{adami} Adami, C, Self-organized criticality in living systems,
     {\it Phys. Lett. A}, {\bf 203}, 23 (1995).

\bibitem{appert} Appert C. and Zaleski S., Lattice gas with a
liquid-gas transition, {\it Phys. Rev. Lett.}, {\bf 64}, 1-4
(1990).

\bibitem{bak87} Bak P, Tang C and Wiesenfeld K, Self-organized
criticality: an explanation of 1/f noise, {\it Phys. Rev. Lett.},
{\bf 59}, 381-384 (1987).

\bibitem{bak96} Bak P, {\it How nature works:the science of
self-organized criticality}, Springer-Verlag, (1996).


\bibitem{boc} Boccara N. and M. Roger,
    Some properties of local and nonlocal site exchange deterministic
    cellular automata, {\it Int. J. Modern Phys.}, {\bf C5},581-588 (1994).

\bibitem{bollobas} Bollobas B., {\it Random graphs}, Academic Press, New York, (1985).

\bibitem{boots} Boots M., Sasaki A., Small worlds and the evolution of virulence:
      infection occurs locally and at a distance, {\it Proc Roy Soc Lond}, B {\bf 266},
      1933-1938 (1999).

\bibitem{arc02} De Arcaneglis, L. and Herrmann, H. J.,
Self-organized criticality on small world networks, {\it Physica
A}, {\bf 308}, 545-549 (2002).


\bibitem{guinot} Guinot V., Modelling using stochastic, finite state
cellular automata: rule inference from continuum model, {\it Appl.
Math. Model.}, {\bf 26}, 701-714(2002).

\bibitem{moukarzel} Moukarzel C. F., Spreading and shortest paths in systems with sparse
long-range connections, {\it Phys. Rev.} E, {\bf 60}, 6263-6266,
(1999).

\bibitem{newman2000} Newman M. E. J., Moore C., Watts D. J., Mean-field
solution of the  small-world network model, {\it Phys. Rev.
Lett.}, {\bf 84}, 3201-3204 (2000).

\bibitem{newman99} Newman M. E. J., Watts D. J., Scaling and percolation
in the small-world network model, {\it Phys. Rev.} E, {\bf
60},7332-7342 (1999).

\bibitem{pandit} Pandit S. A., Amritkar R. E., Characterization and control
of small-world networks, {\it Phys. Rev. E}, {\bf 60}, 1119-1122
(1999).

\bibitem{Tomassini} Tomassini M., {\it Generalized automata networks},
7th Int. Conference on Cellular Automata for Research and Industry,
ACRI 2006, France, {\it Lecture Notes in Computer Sciences}, {\bf
4173}, 14-28 (2006).

\bibitem{Tomas1} Tomassini M., Giacobini M., Darabos C.,
Evolution and dynamics of small-world cellular automata, {\it
Complex Systems}, {\bf 15}, 261-284 (2005).

\bibitem{watts99} Watts D. J., {\it Small worlds: The dynamics of networks
between order and randomness}, Princeton Univ. Press, (1999).

\bibitem{watts98} Watts D. J., Strogatz S. H., Collective dynamics of small-world
networks, {\it Nature} (London), {\bf 393}, 440-442 (1998).

\bibitem{weimar} Weimar J. R., Cellular automata for reaction-diffusion systems,
{\it Parallel computing}, {\bf 23}, 1699-1715 (1997).

\bibitem{wolfram94} Wolfram, S., {\it Cellular automata and complexity},
Reading, Mass: Addison-Wesley (1994).

\bibitem{yang01} Yang X. S., Chaos in small-world networks, {\it
Phys. Rev. E.}, {\bf 63}, 046206 (2001).

\bibitem{yang02} Yang X. S. and Young Y., Cellular automata, PDEs,
and pattern formation, in: Handbooks of Bioinspired Algorithms and
Applications (eds. Olarius S. and Zomaya A. Y.),Chapman \& Hall /CRC
Press,  271-282 (2005).

\bibitem{yang03} Yang X. S., Computational modelling of nonlinear
calcium waves, {\it Applied Maths. Modelling}, {\bf 30}, 200-208
(2006).

\end{thebibliography}
\end{document}